# Charge-carrier lifetime measurements in early-stage photovoltaic materials: intuition, uncertainties, and opportunities

Jeremy R. Poindexter[1], Edward S. Barnard[2], Rachel C. Kurchin[1], Tonio Buonassisi[1]

[1]*Massachusetts Institute of Technology, Cambridge, Massachusetts 02139, USA*
[2]*The Molecular Foundry, Lawrence Berkeley National Laboratory, Berkeley, California 94720, USA*

**Abstract**

Measurements of charge-carrier lifetime in many early-stage thin-film photovoltaic materials can be arduous due to the prevalence of defects and limited information about material properties. In this perspective, we give a brief overview of typical techniques for measuring lifetimes and discuss the intuition involved in estimating lifetimes from such techniques, focusing on time-resolved photoluminescence as an example. We then delve into the underlying assumptions and uncertainties involved in analyzing lifetime measurements. Finally, we outline opportunities for improving accuracy of lifetime measurements by utilizing two emerging techniques to decouple different recombination mechanisms: two-photon spectroscopy, which we demonstrate on BiI$_3$ thin films, and temperature- and injection-dependent current–voltage measurements.

**Introduction**

Efforts to discover and develop new thin-film photovoltaic (PV) absorbers comprising Earth-abundant and non-toxic constituents are accelerating in hopes of lowering PV manufacturing capital costs—which limit industry growth[1,2]—while also resolving scalability and stability concerns with current high-efficiency thin-film technologies (namely CdTe, CuIn$_x$Ga$_{1-x}$Se$_2$, and lead halide perovskites). In evaluating the potential of early-stage PV materials, the charge-carrier lifetime[*] (henceforth simply "lifetime") is crucial to assess since it must exceed at least ~1 ns to enable device efficiencies exceeding 10%[3]—a threshold many of these materials have struggled to reach[4], presumably due to the prevalence of various defects. However, lifetime measurements are not always straightforward in these materials. Standard lifetime techniques exist to inject charge carriers into absorbers and monitor any resulting changes in optical and electrical response (e.g., conductivity and photoluminescence), but many of these techniques are more difficult in thin-film materials, in which lifetimes are typically in the picosecond or nanosecond range, often resulting in low signal-to-noise ratios and detection or sensitivity limits. Furthermore, even if measurements are successful, their analysis can be complicated and confusing due to film heterogeneity at different length scales, the unknown presence of recombination-active defects, lack of information about basic materials properties (e.g., intrinsic charge-carrier recombination coefficients, background carrier concentration, charge-carrier mobilities), and the larger effect of surfaces on recombination due to much larger surface-area-to-volume ratio of thin films compared to thick crystals (e.g., Si wafers).

---

[*]Here, we prefer the term "charge-carrier lifetime" rather than the more traditional "minority-carrier lifetime." Theoretically, one can define separate "minority-carrier" and "majority-carrier" lifetimes, but most lifetime measurement techniques actually observe *recombination* (via the excess carrier concentration), which intrinsically involves both electrons and holes, even if one carrier type dominates. As such, we use "charge-carrier lifetime" to avoid the impression that the lifetime is determined solely by one carrier type.

In this perspective, we give a brief overview of lifetime measurement techniques in semiconductors, focusing on their application to early-stage PV materials. Using time-resolved photoluminescence as an example, we discuss (1) the *intuition* involved in planning experiments for measuring lifetimes; (2) *uncertainties* in such measurements and how to account for them, emphasizing the risk in naïvely conflating lifetimes with PL decay time constants; and (3) *opportunities* for advancing the state of the art of lifetime measurements by decoupling recombination mechanisms using (a) two-photon photoluminescence spectroscopy and (b) temperature- and illumination-dependent current–voltage measurements.

**Guiding intuition for performing charge-carrier lifetime measurements**

*Defining charge-carrier lifetime*

In semiconductors, electrons can be excited to higher energy states (e.g., due to the absorption of a photon), creating electron-hole pairs—a process called *generation*. At some later time, these electrons will fall back to their ground states through *recombination*. The charge-carrier lifetime is typically defined as the average time it takes for electrons and holes to recombine after excitation. Because charge-carrier lifetime is defined as this statistical average—involving large populations of electrons and holes all recombining at different times via multiple pathways—it cannot be measured directly; in other words, it remains prohibitively difficult to track individual charge-carriers, sum up their individual lifetimes, and precisely calculate their average. As such, lifetime must be somehow inferred from another property such as photoluminescence, photoconductance, or photovoltaic response in which generation and recombination both occur.

Rates for both generation ($G$) and bulk recombination ($R$) are incorporated into the continuity equation for charge carriers in semiconductors (shown here for electrons):

$$\frac{\partial (\Delta n)}{\partial t} = G - R + D \frac{\partial^2 (\Delta n)}{\partial x^2} + \mu E \frac{\partial (\Delta n)}{\partial x} \tag{1}$$

where $\Delta n$ is the excess carrier concentration, $D$ is the diffusivity, $\mu$ is charge-carrier mobility, $E$ is electric field, and $x$ and $t$ are dimensions of distance and time, respectively. The third and fourth terms on the right side of equation (1) are commonly referred to as *diffusion* and *drift*, respectively.

Lifetime ($\tau$) relates to equation (1) through the bulk recombination term $R$, which typically includes Shockley-Read-Hall[†] (SRH)[5,6], radiative, and Auger recombination. Surface recombination is typically included in boundary conditions, as shown later in equation (5). Because of the reciprocal relationship between $R$ and $\tau$, and since recombination rates add linearly, it bears emphasizing that the *effective* lifetime of a semiconductor is the harmonic sum of all its contributions:

$$\frac{1}{\tau_{\text{eff}}} = \frac{1}{\tau_{\text{SRH}}} + \frac{1}{\tau_{\text{radiative}}} + \frac{1}{\tau_{\text{Auger}}} + \frac{1}{\tau_{\text{surface}}} \tag{2}$$

Other models that directly relate the various $\tau$'s to $R$ are described in Refs. [5–10].

---

[†]One conventionally used expression for SRH recombination, assuming low injection, is $R = \Delta n / \tau$.

Simple analytic solutions to equation (1)—for example, monoexponential decay of the form $\Delta n \propto \exp\left(-\frac{t}{\tau}\right)$—occur only under specific situations, such as in low injection when one recombination type dominates[11]. Thus, obtaining meaningful estimates for $\tau_{eff}$ often requires making simplifying assumptions or approximations and/or utilizing numerical methods. Depending on the accuracy of these assumptions and materials properties, different methods are useful for different absorber materials.

*Choosing a suitable measurement technique*

Different optoelectronic properties can be utilized to perform lifetime measurements, the most common being (1) photoluminescence, (2) photoconductance, and (3) photovoltaic response. **Figure 1** shows select techniques that utilize these three approaches, plotted roughly based on the measurable values of resistivity and lifetime (abbreviations and references are listed in **Table I**). While we do not intend to provide an exhaustive list of techniques, since that has been done previously[11], the context provided here helps guide the intuition needed to choose a suitable technique, understand assumptions inherent in different methods, and begin applying analysis methods. Note that **Figure 1** includes both time-resolved and steady-state techniques, which involve different assumptions applied to equation (1). Other factors in determining a suitable lifetime-measurement technique include sensitivity to shunting, expected mobility (since some techniques estimate the mobility-lifetime product $\mu\tau$), band gap type (i.e., direct vs. indirect), substrate constraints (e.g., transparent to visible light), and device structure (i.e., whether ohmic contacts need to be applied).

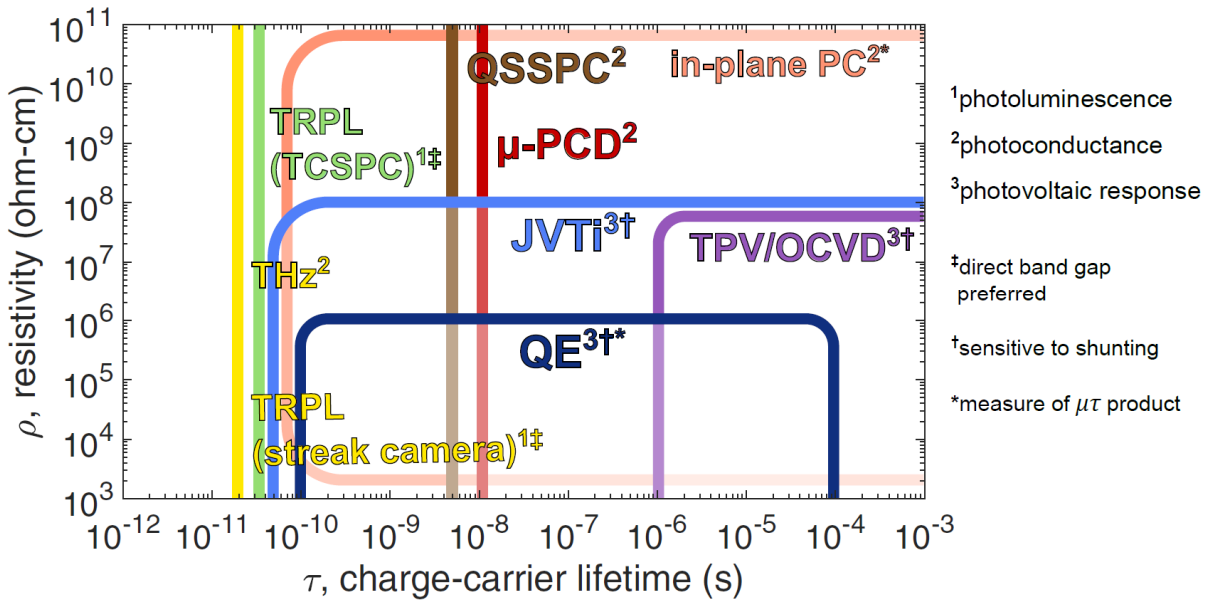

**Figure 1**: Approximate operating ranges of various lifetime measurement techniques depending on film resistivity ($\rho$) and lifetime ($\tau$), categorized by measurement method and other constraints. Abbreviations are listed in Table I.

**Table I**: List of lifetime measurement techniques from Figure 1 with references.

| technique | abbreviation | select references |
|---|---|---|
| time-resolved photoluminescence (via Streak Camera) | TRPL (Streak camera) | 12 |
| time-resolved photoluminescence (via time-correlated single-photon counting) | TRPL (TCSPC) | 13 |
| transient terahertz spectroscopy | THz | 3,14 |
| in-plane photoconductivity | in-plane PC | 11,15 |
| quasi-steady-state photoconductance | QSSPC | 11,16 |
| microwave photoconductivity | μ-PCD | 16 |
| quantum efficiency | QE | 17,18 |
| temperature- and illumination-dependent current–voltage | JVTi | 19 |
| transient photovoltage / open-circuit voltage decay | TPV/OCVD | 11,20 |

*Background on time-resolved photoluminescence*

To serve as an example for guiding intuition, we will focus on (time-resolved) photoluminescence, though its analysis is similar to that for photoconductance since both techniques monitor signals that can be related to $\Delta n(x,t)$, which itself relates directly to solutions to equation (1). Approaches to measure lifetimes from photovoltaic response generally require further assumptions and/or numerical methods; we will highlight these in the "opportunities" section.

Time-resolved photoluminescence (TRPL) has emerged as a useful technique for measuring thin-film PV materials with direct bandgaps in the visible or near-infrared range. Utilizing detectors with time responses of ~50 ps and ultrafast (femtosecond or picosecond) pulsed lasers, one can achieve sub-100-ps resolution in the decay of the photoluminescence (PL) signal after a generation event. Electrical contacts are not needed, and depending on the measurement setup, either optically opaque or transparent substrates can be used. However, weakly emitting samples are problematic, so materials with indirect band gaps can be difficult to measure, especially those with band gaps close to that of the detector (usually Si or InGaAs). A typical setup involves a laser, focusing optics, filters to remove excess laser light, power meter, timing electronics (for single-photon-counting setups), and detector; see Refs. [11,13] for more details.

TRPL measurements are designed to detect photons emitted from the sample via radiative recombination. One should be careful not to then conclude that the timescales involved in the PL decay are determined only by radiative recombination rates. In fact, PL as a function of time relates to the excess carrier concentration as a function of time (and depth within the film), $\Delta n(x,t)$; thus, *all* types of recombination affect the PL dynamics, including Auger and SRH recombination, which are typically non-radiative in nature. It is precisely this fact that allows for the relation of lifetime to PL in the first place. This can be seen in equation (3) below:

$$PL = B(np - n_0 p_0) \tag{3}$$

where $B$ is a material-dependent coefficient for radiative recombination, and $n_0$ and $p_0$ are the equilibrium carrier concentrations for electrons and holes, respectively. Since $np = (n_0 + \Delta n)(p_0 + \Delta p)$, equation (3) can be rewritten as:

$$PL = B(\Delta n \Delta p + p_0 \Delta n + n_0 \Delta p) \tag{4}$$

recognizing that *PL*, $\Delta n$, and $\Delta p$ are all time-dependent terms. Using equation (4), we can now directly relate *PL* to $\Delta n$ from equation (1).

**Uncertainties in interpreting charge-carrier lifetime measurements**

*Injection-level dependence*

In practice, the PL coefficient $B$ is often unknown for early-stage PV materials, and thus is often estimated or included as a fitting parameter[21,22]. The equilibrium carrier concentrations $n_0$ and $p_0$ may also be unknown, although useful upper bounds may be established by the (in)ability to perform Hall effect measurements on films. Therefore, a key uncertainty to understand is the effect of injection-level dependence.

For defect-assisted recombination, the variation of lifetime with injection level can be directly seen from SRH statistics, where $n$ and $p$ affect the overall recombination rate[5,6]. However, whether lifetime increases or decreases with injection level depends on defect parameters—capture cross-section ratios and energy levels, as summarized in Ref. [23]. In materials where there is uncertainty about which defects may be dominating recombination, one cannot make *a priori* assumptions about how lifetime varies with injection. However, some simplifying assumptions can be used in limiting cases.

Under low-injection conditions, equation (4) can be simplified by recognizing that $\Delta n = \Delta p \ll n_0, p_0$. Thus, the expression for PL reduces to $B(p_0 + n_0)\Delta n$ (which may be further approximated by $Bp_0\Delta n$ for a p-type material, for example). In this case, PL $\propto \Delta n$. Under high injection, however, the term $B(\Delta n \Delta p) = B\Delta n^2$ may dominate, in which case PL $\propto \Delta n^2$. These assumptions can have implications on estimated charge-carrier lifetimes, especially in cases where $B$ is unknown. In particular, methods that incorporate full numerical modeling may result in longer lifetimes compared to using exponential fit type models depending on their assumptions about injection[24]. As such, stated assumptions and estimated values can affect results. Additionally, there is often covariance between fitting parameters such as surface recombination and bulk recombination such that finding unique fits can be difficult. Experimentally, varying injection level (i.e., excitation power) and excitation wavelength during measurements can aid in separating out different recombination pathways.

Additionally, because carriers may also diffuse or drift before recombining, PL must account for recombination everywhere within a film by integrating over the film thickness, and recognizing that carriers may propagate laterally. Different models used to perform this integration (e.g., front and back surface recombination vs. semi-infinite solid) can result in different fits[21,25].

*Relating lifetime to recombination rates*

When analyzing time-resolved measurements, ultimately one must either solve equation (1) analytically or obtain a numerical approximation by discretizing $\Delta n$ into units of time and space to determine recombination parameters that comprise $R$, including $\tau_{\text{eff}}$. For systems in which there are too many unknown parameters to obtain unique fits to equation (1), analytical functions such as monoexponential or biexponential fits are used (although expressions including power-law decay[26] and stretched exponentials[27,28] have also been adopted). However, just as radiative recombination rates do not singlehandedly determine the timescales for PL decay, neither does SRH recombination—nor any single recombination mechanism when multiple are present. Thus, one must use extreme caution when applying monoexponential or biexponential fits to data and conflating the resulting timescales with "lifetime." In particular, the "slow" time component of a biexponential decay function, in the context of equation (1), can reflect the emission rate from traps, rather than the lifetime itself[11,29–31]. Additionally, the omission of coefficients of biexponential terms can mislead the reader as to which recombination mechanism actually dominates. A more transparent way of representing the estimated timescales of PL decay would be to apply monoexponential fits separately to different regions of the decay curve.

Alternative formulations to represent lifetimes or recombination can also help avoid confusion. If the purpose of lifetime measurements is to assess general recombination strength rather than identify specific recombination pathways, defining lifetime as the time $n\tau$ required for the PL decay to decrease to $(1/e)^n$ of its initial value (where $n$ is a positive integer, typically 1 or 2) may be a more relevant way to make conclusions about how lifetime varies due to experimental parameters[32]. Similarly, quantifying recombination itself may prove less onerous than quantifying lifetime, and may actually give more insight into charge-carrier dynamics[33]. When using these formulations, one must be clear about definitions, but nonetheless they can be helpful in representing recombination, the actually relevant process in determining solar cell performance.

In **Table II**, we compare approaches to estimating lifetimes using some of these methods. For monoexponential fits, where $PL(t) = A\exp\left(-\frac{t}{\tau}\right)$, we fit different regions of the PL decay from **Figure 2** (see "opportunities" section below) using two fits with separate time constants, $\tau_{\text{fast}}$ and $\tau_{\text{slow}}$. We compare this to a biexponential fit, where $PL(t) = C_1\exp\left(-\frac{t}{\tau_1}\right) + C_2\exp\left(-\frac{t}{\tau_2}\right)$, and separately, the "1/e time" as described above and in Ref. [32], Supporting Information[‡]. Our analysis (performed on BiI$_3$ and CdTe samples; see "opportunities" section for details) suggests that agreement between monoexponential and biexponential fits is susceptible to choice of fitting window, which tends to affect the "1/e time" calculation less. Additionally, **Table II** shows that $\tau_2$ in biexponential fits may exhibit notably small coefficients, which relates to the fact that the PL signal has already decayed significantly—by an order of magnitude or more—after a few nanoseconds. The exception to this is the CdTe 2P signal, whose $\tau_2$ coefficient is much higher, corresponding to the slower PL decay that can be seen visually when observing the data. As such, we emphasize that further clarity about fitting windows and coefficients to exponential fits would better contextualize estimated lifetimes during data analysis.

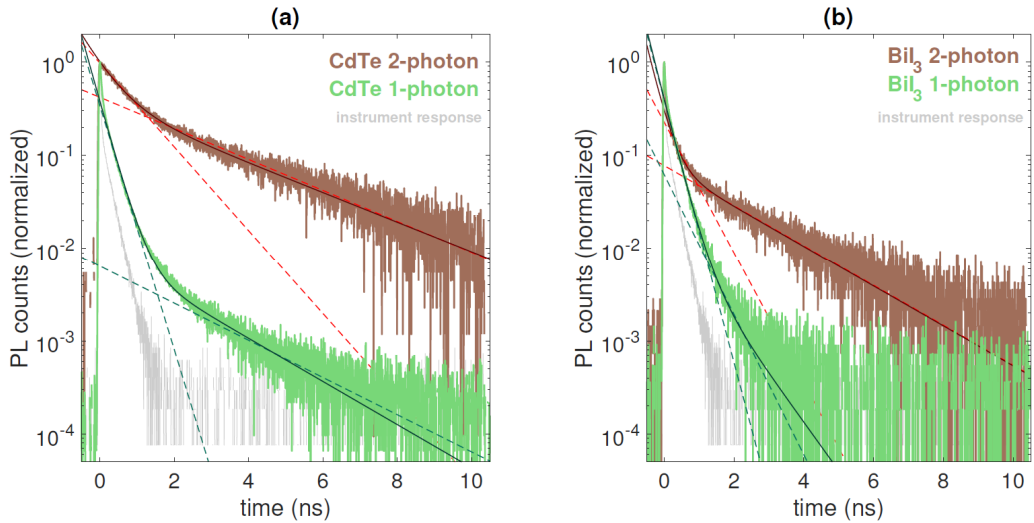

**Figure 2**: TRPL measurements utilizing one-photon (green traces) and two-photon (brown traces) excitation on a CdTe single crystal (a) and a BiI$_3$ thin film (b). The instrument response is shown in gray for comparison.

---

[‡]We use the following equation to calculate two values for $\tau$, for $n$ = 1 ($\tau_{e1}$) and $n$ = 2 ($\tau_{e2}$): $1 - \left(\frac{1}{e}\right)^n = \frac{\int_0^{n\tau} PL(t)dt}{\int_0^{\infty} PL(t)dt}$.

If materials parameters are known with sufficient accuracy, numerical modeling can also be used to fit lifetimes using equation (1), as has been done on various PV materials[24,26,33]. In such cases, fits to experimental data can be insightful, particularly if recombination rates themselves are calculated alongside lifetimes[33]. Such analyses have the benefit of modeling distributions of charge carriers as a function of time and depth within the film, which can reveal insights into where and how recombination may actually occur, perhaps a more meaningful result than the mere magnitude of the lifetime itself.

**Table II**: Calculated PL decay constants from the data in Figure 2 using monoexponential fits, biexponential fits, and time $n\tau$ for the signal to decay to $(1/e)^n$ of its initial value (labeled "$\tau_{en}$", where *n* is an integer).

| sample | mono-exp. $\tau_{fast}$ (ns) | $\tau_{fast}$ fit range (ns) | mono-exp. $\tau_{slow}$ (ns) | $\tau_{slow}$ fit range (ns) | biexp. $\tau_1$ (ns) | biexp. $C_1$ | biexp. $\tau_2$ (ns) | biexp. $C_2$ | biexp. fit range (ns) | $\tau_{e1}$ (ns) | $\tau_{e2}$ (ns) |
|---|---|---|---|---|---|---|---|---|---|---|---|
| **CdTe 1P** | 0.33 | 0.2–1 | 2.17 | 3–8 | 0.30 | 0.37 | 1.87 | 0.009 | 0–8 | 0.28 | 0.38 |
| **CdTe 2P** | 0.97 | 0–1 | 2.61 | 4–7 | 0.56 | 0.64 | 2.73 | 0.36 | 0–10 | 1.94 | 2.28 |
| **BiI$_3$ 1P** | 0.31 | 0.2–1 | 0.58 | 1.5–4 | 0.29 | 0.38 | 0.83 | 0.02 | 0.2–4 | 0.26 | 0.31 |
| **BiI$_3$ 2P** | 0.62 | 0.2–1 | 2.02 | 2–8 | 0.27 | 0.23 | 2.02 | 0.08 | 0.2–10 | 1.27 | 1.74 |
| instrument response | 0.22 | 0.2–1 | – | – | – | – | – | – | – | – | – |

*Surface recombination*

As boundary conditions to equation (1), surface recombination, *S*, can also affect the carrier concentration Δ*n*, and thus the PL as a function of time:

$$D\frac{\partial(\Delta n)}{\partial x}\bigg|_{x=0,d} = S\Delta n \quad (5)$$

Thus, surface recombination can also obscure changes in PL due to recombination, providing additional complications to analysis. In thin-film materials, surfaces tend to dominate much more strongly than in bulk crystals or materials with low absorption coefficients. While surface recombination is typically used as a catch-all term for a defect-rich interface, it can be thought of as a process involving a myriad of SRH-type defects to the point that successive non-radiative recombination steps proceed in sequence. As such, surface recombination can exhibit injection dependence as well, which is beginning to be accounted for in some models[34]. In general, one should be careful when assuming certain regions of PL decay curves are due to surface vs. bulk recombination, as anything that affects Δ*n* can affect the PL, sometimes in unexpected ways[35]. Large values of *S* might be mistaken for "fast" modes of bulk recombination. This type of recombination can also occur internally within multicrystalline films at grain boundaries, but such behavior is harder to capture numerically as it requires the use of 2D or 3D models.

## Opportunities for reducing uncertainties in lifetime measurements

Despite the many uncertainties involved in measuring lifetimes in early-stage PV materials, there remain opportunities for innovation in both measurement and analysis techniques.

*Two-photon spectroscopy*

Employing the use of two-photon absorption, in which a semiconductor is optically excited using sub-bandgap light, offers opportunities for probing sub-surface charge-carrier dynamics. Two-photon absorption is a nonlinear optical effect that only occurs in regions of high photon flux.

Typically, this requires the use of a pulsed laser system. By controlling the location of a high-photon-flux region (e.g., by using a focusing optic), carriers can be generated at arbitrary locations, including below surfaces, while in low-photon-flux regions sub-bandgap light is simply transmitted or reflected. Thus, the typical Beer-Lambert generation profile common in PL measurements can be avoided. Instead, generation profiles that peak *beneath* the surface are possible, and it is therefore possible to decrease or eliminate the effect of surface recombination—which in some cases can dominate $\tau_{eff}$—on lifetime measurements.

We demonstrate this technique by probing CdTe single crystals (borrowing the sample and procedure from Ref. [35]) and $BiI_3$ thin films with both one-photon (532 nm) and two-photon (1064 nm) excitation, then measuring the resulting photoluminescence. TRPL decay curves are shown in **Figure 2**. The two-photon measurements, which in both cases were taken ~2 μm beneath the surface of the film, show much slower PL decay overall in both cases, suggesting that surfaces may limit recombination in both CdTe and $BiI_3$ one-photon measurements. Plane-view maps of the PL intensity (collected separately), shown in **Figure 3**, also demonstrate that two-photon measurements generally are less sensitive to surface features.

However, one must be somewhat careful when making direct comparisons between one- and two-photon measurements, as one- and two-photon excitation can result in much different injection levels. While in theory this can be corrected for by adjusting the excitation power depending on the two-photon absorption coefficient, in practice few two-photon absorption coefficients are known, especially for early-stage PV materials. One workaround is to adjust the excitation power until roughly the same number of PL counts are generated in both one- and two-photon excitation. Furthermore, it is possible that different spin-selection rules for carrier excitation may result in different excited carrier dynamics, although this possibility has been little explored.

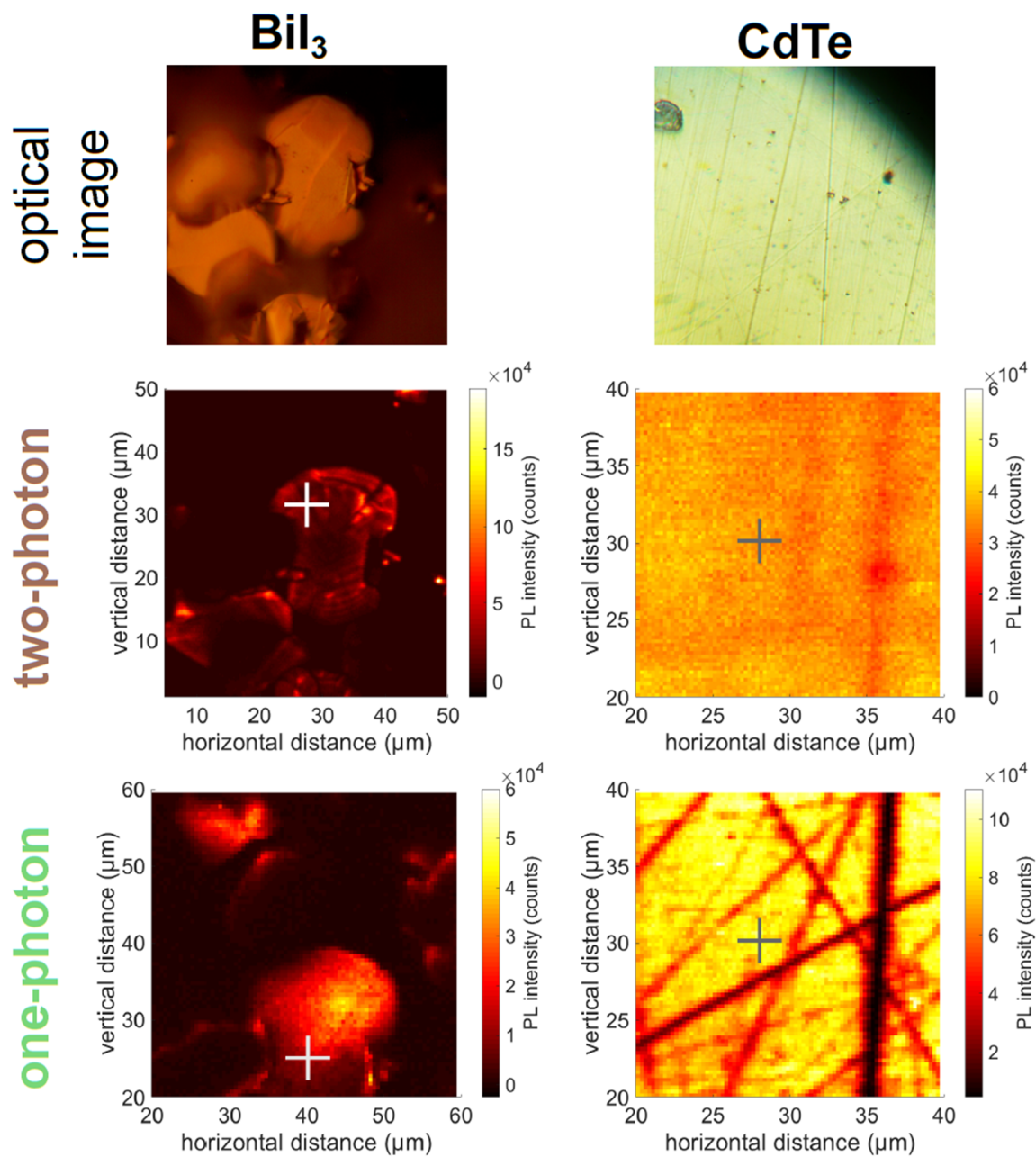

**Figure 3**: Optical microscope images (top row) of CdTe (left column) and BiI$_3$ (right column) samples. PL intensity maps of regions of CdTe and BiI$_3$ samples are also shown for two-photon (middle row) and one-photon (bottom row) excitation. The crosshairs (plus symbols) on the maps indicate the location of measurements from Figure 2, taken approximately 2 μm beneath the surface.

*Advanced PV device modeling techniques*

With sufficient modeling capabilities, lifetime can be extracted or inferred from electrical measurements of full solar cell devices. For example, quantum efficiency measurements—typically used to diagnose current losses—can be modeled to extract mobility-lifetime product $\mu\tau$, and the application of voltage and light biases can assist in this process, though uncertainties (most notably, absorption coefficient) still exist here[17]. Similarly, temperature- and injection-dependent current–voltage measurements can be used to diagnose voltage losses (e.g., by analyzing temperature-dependent $V_{OC}$)[36]. In these cases however, the large number and uncertainties of input parameters can complicate the process of obtaining reasonably narrow confidence intervals for inferred lifetimes. Attempts to use analytic approaches—e.g., double-diode models—remain limited to particular device architectures and are difficult to transfer to structures for early-stage PV materials due to non-idealities in current–voltage behavior.

With the advent of machine learning and advanced statistical approaches, researchers in many fields are using Bayesian-inference-type approaches to infer input parameters from simulated data. Recently, this approach has been applied to solar cells based on tin monosulfide (SnS)[19] to confirm short lifetimes around 50 ps, in agreement with previous THz spectroscopy measurements[3]. Additionally, the Bayesian inference approach was able to infer a surface recombination velocity of 1000–1800 cm/s in SnS devices, a previously unknown parameter[19]. By combining this computational approach to experimental measurements, lifetimes and other unknown device parameters can be more accurately inferred. Such an approach could also be replicated to better model TRPL or photoconductance measurements to reduce uncertainties and increase flexibility in methods for extracting lifetimes, mobilities, and other unknown parameters.

**Outlook and conclusions**

To assess how the process of developing PV materials can be accelerated, we have outlined potential techniques for measuring charge-carrier lifetimes, focusing on the applicability of these techniques to early-stage thin-film materials. Our breakdown of these techniques by magnitude of resistivity and lifetime, band gap character, and sensitivity to shunting should provide researchers with a more comprehensive framework for choosing the technique that is most likely to yield positive results. We discuss the intuition involved in relating lifetimes to experimental measurements (particularly time-resolved photoluminescence), encouraging experimentalists to remain aware of uncertainties due to unknown material properties and use exponential fitting procedures judiciously. Two-photon spectroscopy measurements on $BiI_3$ result in longer extracted PL decay time constants (1–2 ns) compared to typical one-photon excitation (below 1 ns), which demonstrates the potential for decreasing the effect of surface recombination. This and similar techniques, including advanced PV device modeling, offer other ways to reduce uncertainties in lifetime, which we hope will further facilitate the development of early-stage PV materials.

**Experimental methods**

The bismuth triiodide ($BiI_3$) thin film was grown by physical vapor transport of $BiI_3$ source powder (99.999%, Alfa Aesar) onto a sapphire substrate, both placed inside a two-zone quartz tube furnace, as described in Ref. [24]. The furnace was purged three times with nitrogen to remove impurities from the ambient air introduced during loading, then pumped to a base pressure of 16.2 mTorr. During growth, nitrogen gas was flowed through the tube at 10 mL/min to maintain an operating pressure of 5 Torr and a flow from source to substrate. The source powder was heated to 275 °C, while the substrate was kept at 175 °C, both for 90 min before cooling. The source temperature was reduced to room temperature over 10 min, while the substrate temperature was reduced at 2 °C/min.

Both one-photon (532 nm with laser pulse fluence of 1.5 µJ/cm$^2$) and two-photon (1064 nm with laser pulse fluence of 13 µJ/cm$^2$) photoluminescence measurements were performed on BiI$_3$ and CdTe at the Molecular Foundry at Lawrence Berkeley National Laboratory. Pulsed laser light from a Coherent Chameleon Ultra II Compact OPO 150-fs pulsed Ti:sapphire laser with a 12 MHz repetition rate was reflected off a 950 nm dichroic filter, through a NA=0.95 objective, and onto the same location for both one- and two-photon excitation for each sample. Emitted light was then transmitted through the dichroic, through a 945 nm short pass filter (plus 550 nm long pass filter for one-photon excitation), spatially filtered with a 100 µm pinhole, and finally absorbed a single-photon avalanche photodiode detector (MPD PDM-series). A PicoQuant PicoHarp 300 time-correlated single-photon counting module was used to collect time-resolved data. For additional details about the microscope, objective, sample stage, and detector, and CdTe sample, see Ref. [35].

## Acknowledgements

The authors would like to thank Rupak Chakraborty and Riley E. Brandt for helpful discussions about lifetime, as well as P. James Schuck for institutional support at LBNL. Two-photon spectroscopy was performed at the Molecular Foundry was supported by the Office of Science, Office of Basic Energy Sciences, of the U.S. Department of Energy under Contract No. DE-AC02-05CH11231. J. R. P. acknowledges support from the Bay Area PV Consortium, the Martin Family Society of Fellows for Sustainability, and the Switzer Environmental Fellowship. R. C. K. acknowledges support from the Mexican Secretariat of Energy (SENER), the Massachusetts Clean Energy Center, an MIT Energy Initiative Total Graduate Fellowship, and the Blue Waters Graduate Fellowship.

## Author Contributions

J. R. P. wrote the manuscript, while all authors contributed feedback and suggestions. E. S. B. and J. R. P. performed two-photon PL spectroscopy measurements. R. C. K. fabricated the BiI$_3$ sample. T. B. supervised the study.